\documentclass[aps,prb,twocolumn,floats]{revtex4}
\usepackage{epsfig,rotating,latexsym}

\draft

\preprint{Version 29.07.1999}

\begin{document}

%-----------------------------------------------------------------
\title{Magnetization in short-period mesoscopic electron systems}

\author{Vidar Gudmundsson and Sigurdur I.\ Erlingsson}
\address{Science Institute, University of Iceland, 
         Dunhaga 3, IS-107 Reykjavik, Iceland}
\author{Andrei Manolescu}
\address{Institutul Na\c{t}ional de Fizica Materialelor, C.P. MG-7 
         Bucure\c{s}ti-M\u{a}gurele, Rom\^ania}
%
%\maketitle
%
%----------------------------------------------------------------

\begin{abstract} 
We calculate the magnetization of the two-dimensional electron gas in a
short-period lateral superlattice, with the Coulomb interaction included
in Hartree and Hartree-Fock approximations.  We compare the results for
a finite, mesoscopic system modulated by a periodic potential, with the
results for the infinite periodic system.  In addition to the expected
strong exchange effects, the size of the system, the type and the strength
of the lateral modulation leave their fingerprints on the magnetization.
\end{abstract}

\pacs{71.45.Gm,71.70.Di,73.20.Dx}

\maketitle
%----------------------------------------------------------------
%\narrowtext
\section{Introduction}
Several different probes have been used to investigate the properties
of the two-dimensional electron gas (2DEG) in the quantum Hall
regime, transport and optical experiments, or equilibrium methods
including capacitance and magnetization measurements, to name only
few.  The magnetization of high mobility homogeneous 2DEG has been
measured by two methods. One method uses sensitive mechanical, torque
magnetometers.\cite{Eisenstein85:875,Wiegers97:3238}  More recently, a
low-noise superconducting quantum interference device (SQUID) has also
been used.\cite{Grundler98:693,Meinel98a,Meinel99:819} These precision
measurements reveal many-body effects such as the exchange enhancement at
odd filling factors \cite{Ando74:1044} and the fractional quantum Hall
effect and, in addition, an unidentified effect around filling factor
$\nu\approx 2$ that might be connected to skyrmions.\cite{Meinel99:819}

These experimental techniques are expected to be used soon
for measuring the magnetization in lateral superlattices. The
magnetization has been calculated for a disordered homogeneous 2DEG
within the Hartree-Fock approximation (HFA)\cite{MacDonald86:2681}
and within a statistical model for inhomogeneities corresponding to a
Hartree approximation (HA). \cite{Gudmundsson87:8005} The results
show strong exchange effects, already observed in experiments,
\cite{Wiegers97:3238,Meinel98a,Meinel99:819} and manifestations of the
screening properties of the 2DEG.

Concerning periodic systems, the magnetization and the persistent current
have been calculated by Kotlyar et.\ al.\ for a finite array of quantum
dots using a Mott-Hubbard model for the electron-electron interactions,
both intra-dot and inter-dot.\cite{Kotlyar97:R10205,kotlyar98:3989}
In an infinite lateral superlattice, defined by a potential periodic
in two directions (electric modulation) the energy spectrum in the
presence of a magnetic field can only be calculated when the ratio of
the magnetic flux through a lattice cell to the unit flux quantum,
$\phi /\phi_0$, is a rational number.  The unit cell can then be
enlarged to have an integer number of flux quanta flowing through
it.\cite{Hofstadter76:2239} In principle, the magnetization of a
2DEG in a lateral superlattice can be evaluated in the thermodynamic
limit as the negative derivative of the free energy with respect to
the magnetic field $B$,\cite{MacDonald86:2681,Gudmundsson87:8005}
since the energy spectrum is always a continuous function of the flux.
\cite{Hofstadter76:2239,Ketzmeric98:9881,kotlyar98:3989} The inclusion of
the Coulomb interaction to the model, within a self-consistent scheme such
as the HA \cite{Gudmundsson95:16744} or the HFA,\cite{Manolescu99:5426}
does severely limit the possibility to effectively change the value of $B$
by a small amount in order to take a numerical derivative.  Nevertheless,
such a thermodynamic calculation can be done for a modulation that varies
only along one spatial direction, i.\ e.\ for an array of quantum wires.

In this paper we shall evaluate the magnetization of a 2DEG in a periodic
potential corresponding to a weak density modulation in both, or in only
one spatial direction.  In other words our system is either an array
of dots or antidots, or of parallel quantum wires, in most cases with
a strong overlap.  We first consider a finite system with boundaries,
and then the unbound system.

For the finite system we are able to calculate the total magnetization.
For the infinite system with a 2D potential
we shall rather use the definition applicable to a mesoscopic system
with a phase coherence length $L_{\phi}$ much larger than the spatial
period of the square unit cell $L$. It is clear to us that in this
manner we are calculating the contribution to the magnetization due to
the periodic modulation, neglecting the contribution stemming from the
edge of a real system.  In an experiment the total magnetization is indeed
measured and we would have a hard time arguing that the edge contribution
is statistically insignificant in the thermodynamic limit.  Our way out of
this dilemma is to compare several results that we can obtain.

First, we compare the results for the finite system of various sizes,
by heuristically separating the contribution of the bulk and edge
current distributions to the total magnetization.  Second, we compare
quantitatively and qualitatively the bulk contribution in the finite
system to the magnetization produced by one unit cell in the infinite
system.  We point out that experimentally it may be possible to measure
only the contribution to the magnetization caused by the periodic
modulation by placing the entire SQUID-loop well within the sample.  And
third, we compare to the total thermodynamic magnetization of an infinite
system which is periodic in only one spatial direction and thus not
subject to commensurability difficulties.

Our calculations show that the bulk contributions to the magnetization are
strongly dependent on the presence or absence of the exchange interaction
in the models supporting the view of Meinel et.\ al.\ that magnetization
is an ideal probe of the many-body effects in a 2DEG.\cite{Meinel99:819}
Magnetization has been calculated for quantum Hall systems
in the fractional regime with higher order approximations
that reproduce more reliably exchange and correlation
effects.\cite{Haussmann97:9684,MacDonald98:R10171} Here we focus our
attention on relatively large structured systems with several Landau
bands included and have thus to resort to the HFA in order to make the
calculation tractable in CPU-time. Recent comparison between results
from exact numerical diagonalization and the HFA show that
in high Landau levels the two approaches agree on the formation
of charge density waves.\cite{Rezayi99:03258}

We shall consider periodic potentials of a short period, 50 nm, which
means short with respect to the present technical possibilities, but still
realistic.  In this case, for GaAs parameters and for magnetic fields
in the range of few Tesla, the screening effects due to the direct Columb
interaction are weak.  However, the exchange effects remain strong, and
they amplify the single-particle energy dispersion.\cite{Manolescu99:5426}
Therefore the presence of the periodic potential should become proeminent
in the magnetization even for weak amplitudes.  And last, but not least,
by avoiding strong screening we are benefitted by a shorter 
computational time.

%
%---------------------------------------------------------------- 
%
\section{Models}
The magnetization is calculated within three models in the paper,
self-consistently with respect to the electron-electron Coulomb
interaction: A finite model using the HA, an infinite model periodic in
both spatial directions, using the unrestricted Hartree Fock approximation
(UHFA), and an infinite model periodic in only one spatial direction,
using the standard HFA.
\subsection{Finite 2DEG}
The model for a finite system consists of a laterally confined 2DEG.  A hard 
wall potential ensures that the electrons stay in the square region 
\begin{equation}
      \Sigma=\left \{ (x,y)|\, 0<x<L_x, 0<y<L_y \right \} \,,
\end{equation}
the wave functions being zero at the boundary.  An external modulating
potential and a perpendicular magnetic field are applied.  The potential
has the form
\begin{equation}
      V_{\mbox{\scriptsize sq}}({\mathbf r})=V_0 \left \{
                           \sin \left ( \frac{n_x \pi x}{L_x} \right )   \
                           \sin \left ( \frac{n_y \pi y}{L_y} \right )   \
                               \right \}^2 \,,
\label{V_sq}
\end{equation}
where $n_x$ and $n_y$ count the number of dots in $x$ and $y$ direction
respectively, giving in total $N_c=n_x n_y$ unit cells.
The Schr{\"o}dinger equation is solved by expanding the eigenfunctions
in Fourier sine-series and the expansion coefficients are found by
diagonalizing the Hamiltonian matrix.  The electron interaction is taken
into account in the Hartree approximation. 

The total magnetization can be calculated according to the definition for
the orbital and the spin component of the magnetization,\cite{Desbois98}
\begin{eqnarray}
      M_o+M_s&=&\frac{1}{2c{\cal A}}\int_{\cal A} d^2r \left( {\bf r}\times
      \langle {\bf J}({\bf r}) \rangle \right) \cdot\hat{{\bf n}}
      \nonumber\\
      &-&\frac{g\mu_B}{{\cal A}}\int_{\cal A} d^2r \langle \sigma_z
      ({\bf r})\rangle \,,  
\label{M_OS}
\end{eqnarray}
where ${\cal A}$ is the total area of the system.
The equilibrium local current is evaluated as the quantum
thermal average of the current operator,
\begin{equation}
      \hat{{\bf J}}=-\frac{e}{2}\bigg(\,\hat{{\bf v}}|{\bf r} 
      \rangle\langle{\bf r}| 
      +|{\bf r}\rangle\langle{\bf r}|\hat{{\bf v}}\,\bigg) ,
\label{J_op}
\end{equation}
with velocity operator 
$\hat{\bf v}=[\hat{{\bf p}}+(e/c){\bf A}({\bf r})]/m^*$,
{\bf A} being the vector potential.
Even though the magnetic field $B$ can be varied freely in this model
we have used the definition of the orbital magnetization (\ref{M_OS})
rather than evaluating the derivative of the free energy with respect
to the magnetic field.

\subsection{Periodic 2DEG in two directions}
The two-dimensional modulation of the system is a  square lattice of 
quantum dots ($V_0>0$) or antidots ($V_0<0$) determined by the static 
external potential
\begin{equation}
      V_{\mbox{\scriptsize QAD}}({\bf r})=V_0
      \left\{ \sin\left(\frac{gx}{2}\right)
      \sin\left(\frac{gy}{2}\right) \right\} ^2,
\label{V_QAD} 
\end{equation}
or a simple cosine-modulation defined by
\begin{equation}
      V_{\mbox{\scriptsize per}}({\bf r})=V_0\left\{\cos\left({gx}\right) 
                               +\cos\left({gy}\right)\right\},
\label{V_per}  
\end{equation}
where $g$ is the length of the fundamental inverse lattice vectors, ${\bf
g}_1=2\pi \hat{\bf x}/L$, and ${\bf g}_2=2\pi \hat{\bf y}/L$.  The Bravais
lattice defined by $V_{\mbox{\scriptsize QAD}}$ or $V_{\mbox{\scriptsize per}}$
has a periodic length
$L$ and the inverse lattice is spanned by ${\bf G}=G_1{\bf g}_1+G_2{\bf
g}_2$, with $G_1,G_2\in {\bf Z}$ . The commensurability condition between the
magnetic length $\ell$ and the period $L$ requires magnetic-field values
of the form $B=pq\phi_0/L^2$, with $pq\in {\bf N}$, and $\phi_0=hc/e$
the magnetic flux quantum.\cite{Silberbauer92:7355,Gudmundsson95:16744}
Arbitrary rational values can, in principle, be obtained by resizing
the unit cell in the Bravais lattice.

For this model we evaluate the contribution of the periodic modulation
to $M_o$ and $M_s$, rather than the total magnetization.  Using in Eq.\
(\ref{M_OS}) the periodicity of the current and spin densities, and the
reflection symmetry of the unit cell, we reduce the integrations to a
single cell.  Obviously, in the absence of the modulation $\langle
{\bf J}({\bf r}) \rangle\equiv 0$ and the orbital contribution vanishes.

The ground-state properties of the interacting 2DEG in a perpendicular
homogeneous magnetic field ${\bf B}=B\hat{\bf z}$ and the periodic
potential are calculated within the UHFA for the Coulomb interacting
electrons at a finite temperature.\cite{Gross91,Palacios94:5760} The
approximation is unrestricted in the sense that the single-electron
states do not have to be eigenstates of $\hat{\sigma}_z$.

\subsection{Periodic 2DEG in one direction}
The modulation is defined by the potential
\begin{equation}
 V_{\mbox{\scriptsize per}}(x)=V_0\cos\left(\frac{2\pi x}{L}
 \right) \,,
\end{equation}
describing an array of parallel quantum wires.  In this case there is no
restriction on the magnetic-field values, the magnetic flux through one
lattice cell always being infinite.  The groundstate is calculated in the HFA,
by diagonalizing the Hamiltonian in the Landau basis, and by expanding
the matrix elements in Fourier series.  Therefore we can evaluate directly 
the total magnetization $M=M_o+M_s$ by the thermodynamic formula 
appropriate for the canonical ensemble,
\begin{equation}
M=-\frac{1}{{\cal A}}\frac{d}{dB}(E-TS),
\label{thermodeq}
\end{equation}
where $E$ is the total energy, and $S$ the entropy.  We shall assume
the temperature is sufficiently low to neglect the second term
of Eq.\ (\ref{thermodeq}).  In view of more realistic results
we also assume a small disorder broadening of the Landau levels,
which we take into account with a Gaussian model for the spectral
function.\cite{MacDonald86:2681}
%
%----------------------------------------------------------------
%
\section{Results}
The numerical calculations are performed with GaAs parameters,
$m^*=0.067$, and $\kappa =12.4$. In the case of the infinite periodic
modulation, in the UHFA, HFA, or the HA, the bare $g$-factor is -0.44,
and is set equal to zero in the model of the finite 2DEG in the HA.
Mostly for numerical reasons we keep a finite temperature, which for the
models with 2D potential is 1 K, and for the 1D potential is 0.2 K. In
all cases the length of the unit cell is $L=50$ nm.

\subsection{Finite system with 2D potential}

\begin{figure} 
\epsfxsize 8cm 
\begin{center}
\epsffile{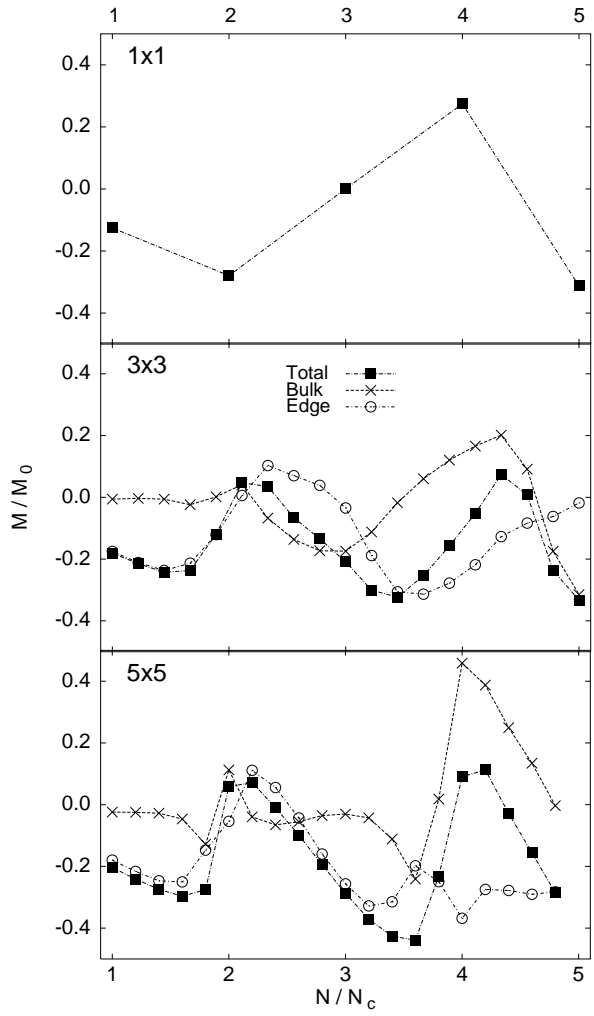}
\end{center}
\caption{The total (orbital) $M_o$, the bulk $M_b$, and the edge $M_e$
         magnetizations of a system of $N$ electrons in
         $n_x\times n_y=N_c$ unit cells (HA).
         $pq$=1, $B\approx 1.65$ T, $M_0=\mu^*_B/(L_xL_y)$,
         $V_0=-1$ meV.}
\label{mag_tot_SIE}
\end{figure}

\begin{figure} 
\epsfxsize 7.4cm 
\begin{center}
\epsffile{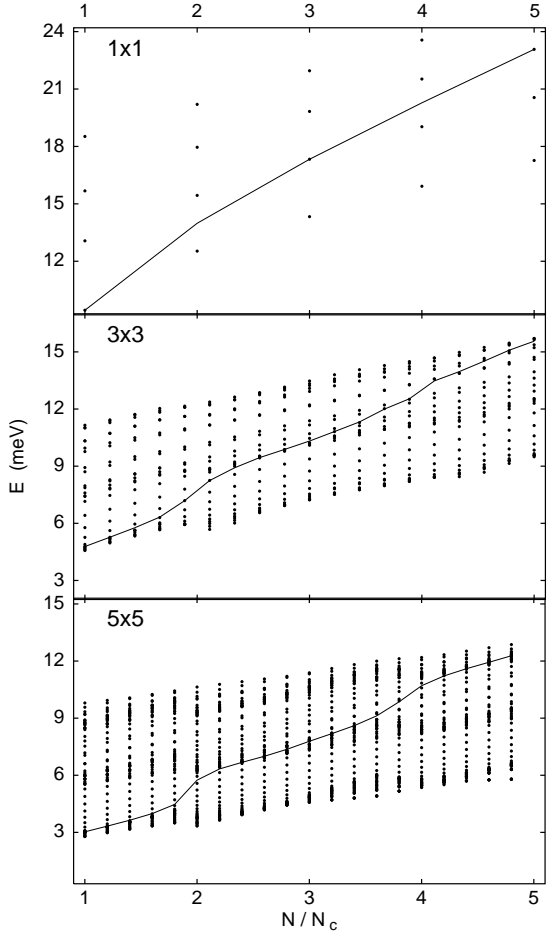}
\end{center}
\caption{The energy spectra (dots) for $N$ electrons 
         and the chemical potential $\mu$
         (solid) for $n_x\times n_y=N_c$ arrays of quantum dots, (HA).
         $pq$=1, $B\approx 1.65$ T, 
         $V_0=\pm 5$ meV.}
\label{EHNs_tot_SIE}
\end{figure}

\begin{figure} 
\epsfxsize 6cm 
\begin{center}
\begin{turn}{-90}
\epsffile{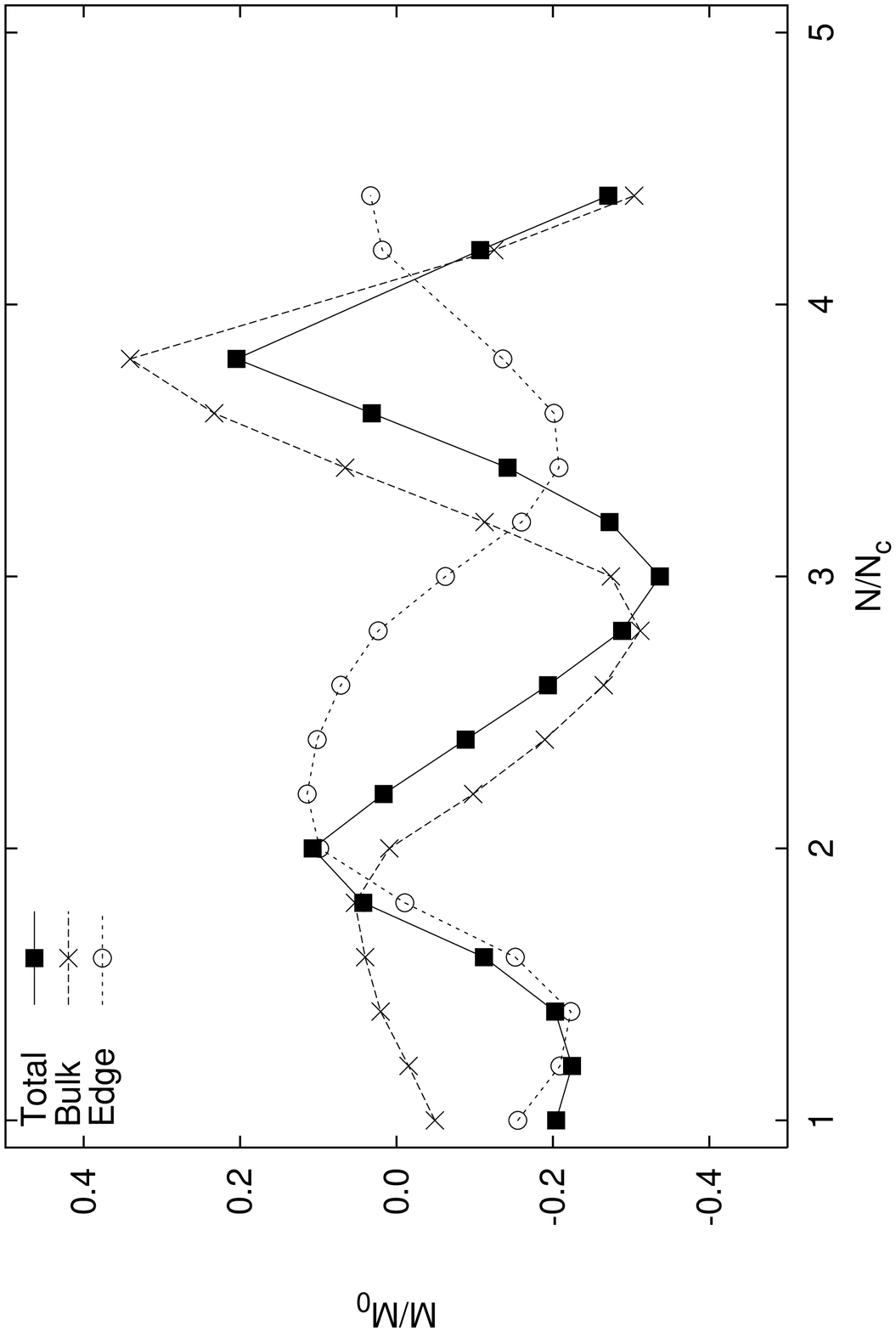}
\end{turn}
\end{center}
\caption{The total (orbital) $M_o$, the bulk $M_b$, and the edge $M_e$
         magnetizations of a noninteracting system of $N$ electrons in
         $5\times 5$ unit cells (HA).
         $pq$=1, $B\approx 1.65$ T, $M_0=\mu^*_B/(L_xL_y)$,
         $V_0=-1$ meV.}
\label{mag_tot_0_SIE}
\end{figure}

The magnetization for the finite system is shown in Fig.\
\ref{mag_tot_SIE}.  The system size is progressively increased, starting
from a single cell of 50$\times$50 nm$^2$, to a system of 5$\times$5
cells, keeping the unit-cell size constant.  Each cell is defined by
one period of the modulation potential (\ref{V_sq}).  When the system
consists of more than one cell we, ad hoc, divide the magnetization
into an edge part $M_e$ with a contribution only from the first row of
cells around the system, and a bulk part $M_b$ with the contribution
from the rest of the cells.  Below we shall see that generally, the
magnetization $M_b$ does approach the orbital magnetization expected for
a large system as the number of cells $N_c$ is increased.  The variable
on the $x$-axis of the figure, $N/N_c$, i.\ e.\ the number of electrons
in a single cell, can approximately be interpreted as a filling factor
for the 3$\times$3 and the 5$\times$5 system. This is confirmed by the
evolution of the chemical potential $\mu$ through the single electron
Hartree-states depicted in Fig.\ \ref{EHNs_tot_SIE}. For even-integer
values of $N/N_c$ $\mu$ jumps through 'gaps' of sparsely distributed edge
states separating states concentrated into precursors of Landau bands. The
bulk $M_b$ and the edge $M_e$ contributions to the magnetization as
well as the total magnetization are of similar magnitudes.  However the
oscillations of $M_b$ around zero are more symmetric than those of $M_e$.
This is because the direction of the edge current is more commonly as
expected from the classical clockwise motion of the electrons around
the sample, thus giving a preferred sign to $M_e$.
Modestly increasing the size, from 3$\times$3 to 5$\times$5 unit
cells, clearly gives the finite system more of the character of
an extended system. The bulk magnetization for the large system
is small when $\nu$ is not close to even integers due to the 
strong screening of the modulation potential away from the edges of 
the system. This is confirmed by Fig.\ \ref{mag_tot_0_SIE} showing
$M_o$ of the noninteracting 5$\times$5 system. The structures around 
even integer values of $\nu$ are less steep than for the interacting 
system. In the presence of the interaction the energy gaps are
reduced by screening, which is self-consistently determined by the
density of states around $\mu$. When $\mu$ lies within 
one 'band' (i.\ e.\ $\nu$ is not an even integer) the 
Coulomb repulsion forces the electron density
to spread out more evenly shifting the ''effective filling``
$N/N_c$ a small amount. A slight change in the magnetic flux in
the finite system only shifts the relation between the number of
electrons and the effective filling factor.\cite{Gudmundsson87:8005}
 
\begin{figure} 
\epsfxsize 6cm 
\begin{center}
\epsffile{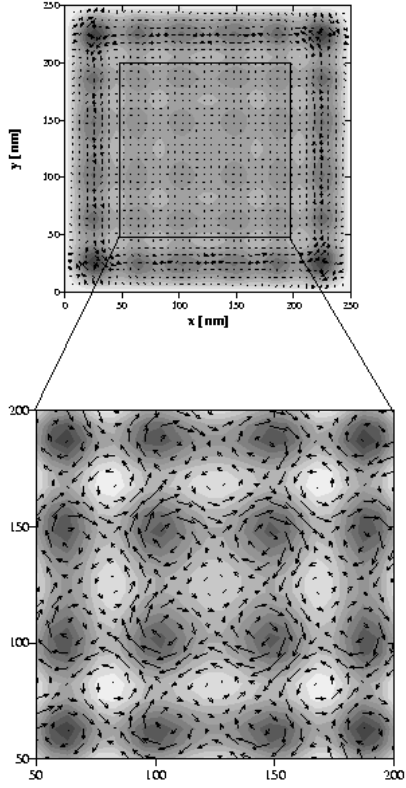}
\end{center}
\caption{The current (arrows) and electron density (contour) 
         in a finite system of 
         5$\times$5 unit cells (HA) for $N/N_c=2.2$ (filling factor).
         $pq$=1, $B\approx 1.65$ T,
         $V_0=-5$ meV.}
\label{curr_SIE}
\end{figure}

The current distribution for a 5$\times$5 system
is shown in Fig.\ \ref{curr_SIE} which reveals a strong edge current,
but also a bulk current structured self-consistently in a complex way by
the modulation, the interaction, and the location of $\mu$ with respect
to the energy levels. This interplay of complex bulk contributions with
the effects of the edge currents opens the question what are the effects
of a modulation in an extended electron system on the magnetization.

\subsection{Infinite system with 2D potential}

\begin{figure} 
\epsfxsize 6cm 
\begin{center}
\begin{turn}{-90}
\epsffile{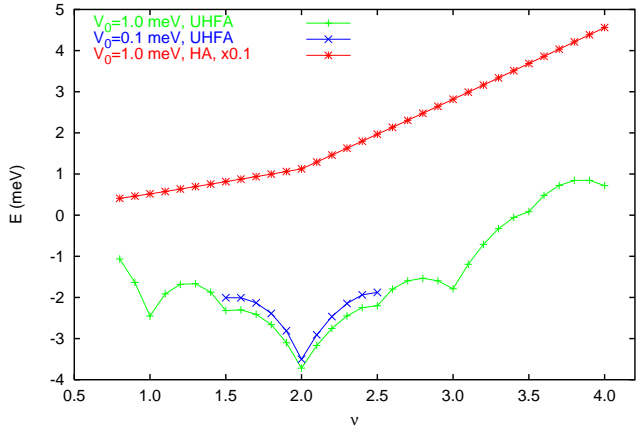}
\end{turn}
\end{center}
\caption{The total energy per unit cell of the electron system 
in the case of the
         simple cosine-modulation (\protect\ref{V_per}) as a function
         of the filling factor $\nu$ for the UHFA and HA.
         ($+$) $V_0=1.0$ meV, and ({\sf x}) $V_0=0.1$ meV.
         In case of the HA the total energy is multiplied by 0.1.  
         $pq$=2, $B\approx 3.3$ T.}
\label{E_tot_cos_pq2}
\end{figure}

\begin{figure} 
\epsfxsize 6cm 
\begin{center}
\begin{turn}{-90}
\epsffile{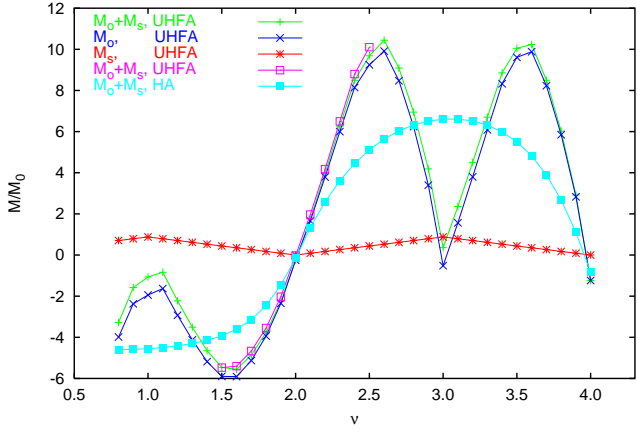}
\end{turn}
\end{center}
\caption{The orbital $M_o$, and spin $M_s$ contribution to the 
         total magnetization for the simple cosine-modulation (UHFA) 
         (\protect\ref{V_per}) with
         ($+$) $V_0=1.0$ meV, and ($\Box$) $V_0=0.1$meV, 
         and the total magnetization in the HA
         for $V_0=1.0$ meV,
         as a function of the filling factor $\nu$.
         $pq$=2, $B\approx 3.3$ T, $M_0=\mu^*_B/(L_xL_y)$.}
\label{M_cos_pq2}
\end{figure}

Next, we turn our attention to the infinite system, that is modulated
in two directions, and calculate the contribution to the magnetization
from one unit cell.  In Fig.\ \ref{E_tot_cos_pq2}  the total energy
is shown as a function of the filling factor $\nu$ for the extended
periodic 2DEG in the UHFA and the HA for the case $pq=2$. Two magnetic
flux quanta flow through the unit cell and each Landau band is split
into two subbands which in turn are doubly spin split.  Filling factor
two means thus that both spin states of one Landau band are occupied,
and in total four subbands are below the Fermi level. The modulation
with $V_{0}=1.0$ or 0.1 meV is small compared to $\hbar\omega_c=5.71$
meV.  The minima in the total energy for the UHFA reflect the strong
exchange interaction for electrons, added to nearly filled Landau
bands or subbands thereof.  Fig.\ \ref{M_cos_pq2}  compares the total
magnetization, Eq.\ (\ref{M_OS}), and its components $M_o$ and $M_s$
calculated within the UHFA, with the total magnetization according to the
HA. The main difference between the results of these two approximations
is the sharp reduction in the magnetization caused by the exchange
interaction around odd integer filling factors. In this region the
enhanced spin splitting of the subbands is larger than the subband
splitting caused by the modulation. The order of the subbands 
(with respect to spin and magnetic subband index) and their
curvature thus leads to $M_o$ being of same sign for $\nu =2.5$ and $\nu
=3.5$. The behavior is thus different around the even and odd values of
$\nu$. Later we see that this is not in the case of $pq=1$. Just like
for the total energy the different modulation strengths result in minor
changes in $M_o$, because the energy dispersion of the Landau bands
is determined by the exchange energy rather than by the external 
or screened potentials. \cite{Manolescu99:5426}  

\begin{figure} 
\epsfxsize 6cm 
\begin{center}
\begin{turn}{-90}
\epsffile{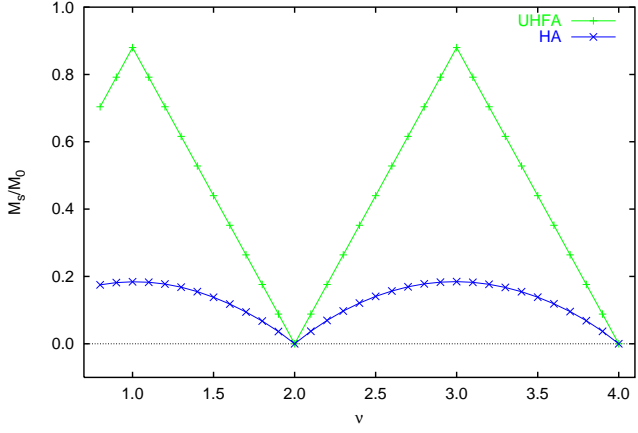}
\end{turn}
\end{center}
\caption{The spin contribution to the magnetization $M_s$ 
         of the electron 
         system in the case of the
         simple cosine-modulation (\protect\ref{V_per}) 
         for the UHFA and the HA as a function
         of the filling factor $\nu$. 
         $pq$=2, $B\approx 3.3$ T, 
         $V_0=1.0$ meV.}
\label{S_mu_pq2}
\end{figure}

The light effective mass and the small $g$-factor of electrons in GaAs
cause the spin contribution $M_s$ to be an order of magnitude smaller
than the orbital one. A comparison of $M_s$ in these two approximations
can be seen in Fig.\ \ref{S_mu_pq2}. In the case of the UHFA the exchange
interaction always leads to the maximum spin polarization. In the HA the
spin splitting of the Landau bands is only the bare Zeeman gap, here about
0.07 meV, i.\ e.\  much smaller than the intra-band energy dispersion which
is of the order of the modulation amplitude, $V_0=1$ meV.  Therefore the
chemical potential is never able to lie only in one spin subband. New
electrons are being added to the system concurrently, to both spin states,
resulting in a reduced polarization.

\begin{figure} 
\epsfxsize 6cm 
\begin{center}
\begin{turn}{-90}
\epsffile{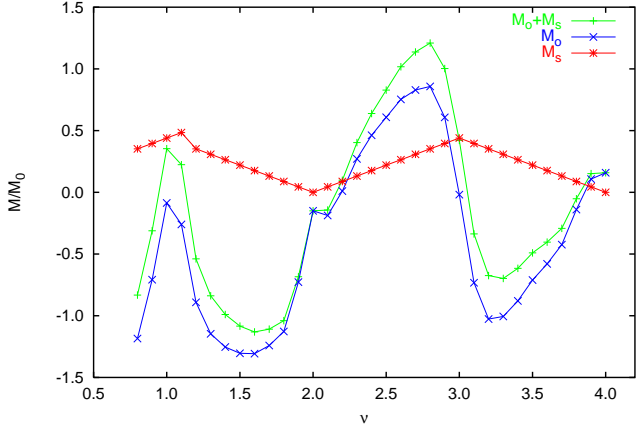}
\end{turn}
\end{center}
\caption{The orbital $M_o$, and spin $M_s$ contribution to the 
         magnetization for the simple cosine-modulation 
         (\protect\ref{V_per}) 
         as a function of the filling factor $\nu$.
         UHFA, $pq$=1, $B\approx 1.65$ T, 
         $V_0=0.1$ meV.}
\label{M_V01_pq1}
\end{figure}

\begin{figure} 
\epsfxsize 8.6cm 
\begin{center}
\epsffile{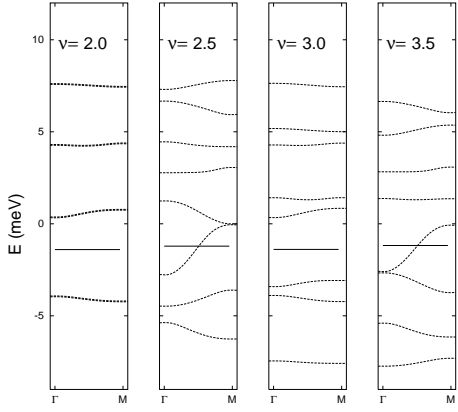}
\end{center}
\caption{Sections ($\Gamma$ - M) of the energy spectra 
         for the dot modulated interacting 2DEG. 
         The chemical potential
         is indicated by a horizontal solid line.
         UHFA, $pq$=1, $B\approx 1.65$ T, $V_0=-1.0$ meV. }
\label{Ebpq1}
\end{figure}

In the case of one flux quantum through a unit cell ($pq=1$) each Landau
band consists of only two subbands, with different spin quantum numbers.
This simpler band structure is reflected in the magnetization (see Fig.\
\ref{M_V01_pq1}). Here the spin-bands hosting $\mu$ for the filling
range $2.5\leq\nu <3$ and $3<\nu\leq 3.5$ have opposite 
curvature causing a maximum and a minimum in $M_o$, respectively. 
For the lower $\nu$ range the Fermi level ($\mu$) is in the lower spin 
subband of the second Landau-band pair, see Fig.\ \ref{Ebpq1}.  
Here the Fermi contours (''Fermi surface``) encircle
the energy maxima at the edges of the Brillouin-zone (hole orbits),
while for the higher $\nu$ range they close around the 
minimum of the upper spin subband (electron orbits). 

\begin{figure} 
\epsfxsize 8.6cm 
\begin{center}
\epsffile{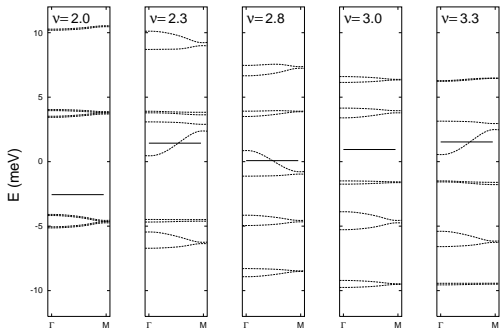}
\end{center}
\caption{Sections ($\Gamma$ - M) of the energy spectra 
         for the simple cosine modulated interacting 2DEG. 
         The chemical potential
         is indicated by a horizontal solid line. 
         UHFA, $pq$=2, $B\approx 3.3$ T, $V_0=1.0$ meV. }
\label{Ebpq2}
\end{figure}

\begin{figure} 
\epsfxsize 7cm 
\begin{center}
\epsffile{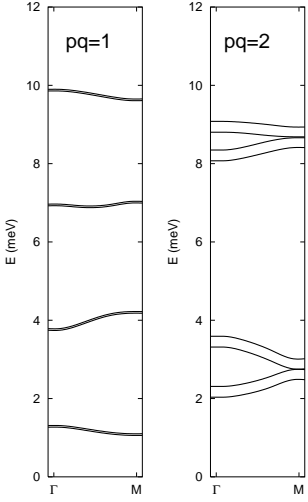}
\end{center}
\caption{Sections of the noninteracting energy spectra corresponding to the
         two interacting cases in Figures 
         {\protect \ref{Ebpq1} and \ref{Ebpq2}}.  
         The small spin splitting is not visible for the
         lower magnetic flux, $pq=1$.}
\label{Eb0}
\end{figure}

In the case of two flux quanta each Landau band is fourfold split
as can be seen in Fig.\ \ref{Ebpq2}. For $\nu =2.0$ only the splitting
due to the modulation is clearly visible but for $\nu$ not equal
to an even integer the spin splitting gets more enhanched due
to the strong exchange force. Furthermore, we see that the subbands
repel each other around the $\Gamma$ point (center of the Brillouin
zone) resulting in a different occurrence as $\nu$ is swept from 2.8 to
to 3.3. In contrast to the strong $\nu$-dependent interaction effects 
on the energy spectra we show in Fig.\ \ref{Eb0} the ''static`` 
energy bands of the noninteracting system that are independent
of $\nu$ and the location of $\mu$. 

Around $\nu=1$ the order of the Landau subbands is unusual.  The states
below the Fermi level are those of the Landau subband $(n,\sigma)=(0,+)$,
$n$ being the orbital and $\sigma$ the spin quantum numbers.  But the
Landau subbands above, i.\ e.\ $(0,-)$ and $(1,+)$, are overlapped due to
the strong exchange enhancement of the gaps between all the subbands
with the same $n$ but opposite $\sigma$.  Therefore the first states
which are populated for $\nu>1$ belong to the subband $(1,+)$, while
the subband $(0,-)$ is populated a bit later (for slightly higher $\nu$).  
This situation
is well known in the atomic physics of complex atoms, and has also been
demonstrated in quantum dots, with a current-spin density-functional
approach,\cite{Steffens98:529} which in principle is more reliable
than our UHFA.  In our case we observe this spin-state inversion as a
small anomaly in $M_s$ around $\nu=1$, Fig.\ \ref{M_V01_pq1}, where the
maximum of $M_s$ is shifted to $\nu>1$.

\begin{figure} 
\epsfxsize 6cm 
\begin{center}
\begin{turn}{-90}
\epsffile{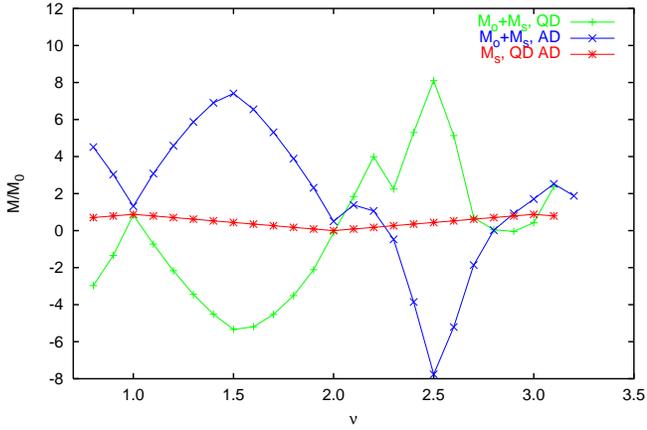}
\end{turn}
\end{center}
\caption{The orbital $M_o$, and spin $M_s$ contribution to the 
         magnetization for a quantum dot (QD), and an antidot 
         (AD) array (\protect\ref{V_QAD}) 
         as a function of the filling factor $\nu$.
         UHFA, $pq$=2, $B\approx 3.3$ T, 
         $V_0=\pm 5$ meV.}
\label{M_V5_QAD_pq2}
\end{figure}

The fact that the magnetization calculated by Eq.\ (\ref{M_OS}) for a
unit cell in an infinite doubly modulated system
reflects {\sl only} the contribution of
the modulation is seen in Fig.\ \ref{M_V5_QAD_pq2}, where $M$ for a dot
and an antidot modulation of the same strength are mirror symmetric around
zero for low $\nu$. For higher $\nu$ the mixing of the Landau bands due to
the Coulomb interaction slightly skews the mirror symmetry. 
For a homogeneous system ($V_0=0$) the persistent current in the  
definition of the magnetization (\ref{M_OS}) vanishes and thus
$M_o$. Similar effect was seen for the bulk magnetization
of the finite system in Fig.\ \ref{mag_tot_SIE} for
noninteger values of $\nu$.  
Due to the enhancement of subband width caused by the 
exchange force the transition to zero magnetization with decreasing
modulation does not happen in a smooth 
linear fashion.\cite{Manolescu99:5426}   

\subsection{Correspondences between the finite and the infinite
systems}

\begin{figure} 
\epsfxsize 6cm 
\begin{center}
\begin{turn}{-90}
\epsffile{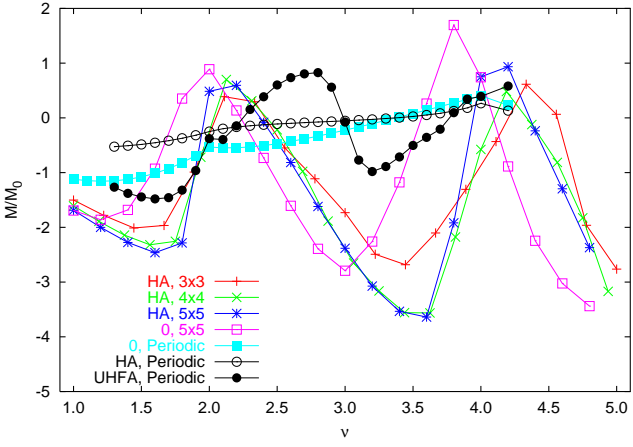}
\end{turn}
\end{center}
\caption{The orbital magnetization $M_o$ for the dot array
         (\protect\ref{V_QAD}) in UHFA, HA, and for noninteracting
         electrons compared to $M_o$ for a finite systems of $n\times n$
         unit cells in HA
         as a function of the filling factor $\nu$.
         (For the finite system $\nu$ is approximated).
         $pq$=1, $B\approx 1.65$ T, 
         $V_0=-1$ meV.}
\label{M_samanb_1H}
\end{figure}

Now we come back to the question how the magnetization of the
finite and infinite system are related. The magnetization for 
the finite system can be calculated by either equation (\ref{M_OS})
or (\ref{thermodeq}) giving the same results for the low temperature
assumed here. The orbital
magnetization $M_o$ is compared for the finite system of different sizes
to the result for a single unit cell in an extended system with the same
type of modulation in Fig.\ \ref{M_samanb_1H}, but with the
interaction accounted for in different ways. If we first look at the
results for the infinite periodic system we notice that the largest 
variation of $M_0$ occurs for the UHFA and the smallest for the HA
with the noninteracting case inbetween. This is in accordance with 
the screening properties mentioned earlier, in the HA the modulation
is screened more effectively than in the UHFA. This result is in
agreement with the simplistically defined bulk magnetization 
$M_b$ of the finite system seen in Fig.\ \ref{mag_tot_SIE}, the
main difference being the sharp variation of $M_b$ at even 
integer values of $\nu$. Their presence indicates that even though 
the current density is only integrated over the ''bulk`` of the
system the underlying energy spectrum is affected by the chemical potential
$\mu$ traversing it's edge states. To be able to get the unit cell     
of the two different systems to give the same magnetization the
finite system has to be even larger, exhausting our means of
computation. 

The magnetization of both systems (and also of the 1D modulated system,
see the next subsection) compares well with results derived from an older
model of statistical inhomogeneities in a 2DEG that was used to explain
effects caused by oscillating Landau level width due to the electrostatic
screening.\cite{Gudmundsson87:8005} 

The main differences between the magnetization of our finite and
infinite periodic systems are:   
{\it i}) the asymmetry around the zero line in the case of the finite system, 
and {\it ii}) the missing steepness of $M_o$ around even integer
filling factors $\nu$ for the infinite system. 
Earlier we saw that the asymmetry is influenced by the contribution
from the ''edge`` of the finite system. The second difference can also
be traced to the edge states.  In the extended model there are no edge
states between the Landau subbands. Their shape and curvature can thus
change sharply with the motion of the chemical potential $\mu$ through
them, the self-consistent screening and exchange effects minimize any
gaps that might evolve around $\mu$, which in turn prevents any sharp
jumps in $M_o$. With this in mind it is clear that {\sl the magnetization
for a realistic (large, but finite) modulated system is not simply the sum of 
the magnetization produced by two independent subsystems, the edge and
the bulk}. The Coulomb interaction makes the separation of the
contributions to $M$ a nontrivial problem, which can be solved only 
by an experiment. In addition, we have seen that
the self-consistent motion of $\mu$ through the energy bands depends
strongly on the approximation used for the electron interaction.

\subsection{Infinite system with 1D potential}

\begin{figure} 
\epsfxsize 8cm 
\begin{center}
\epsffile{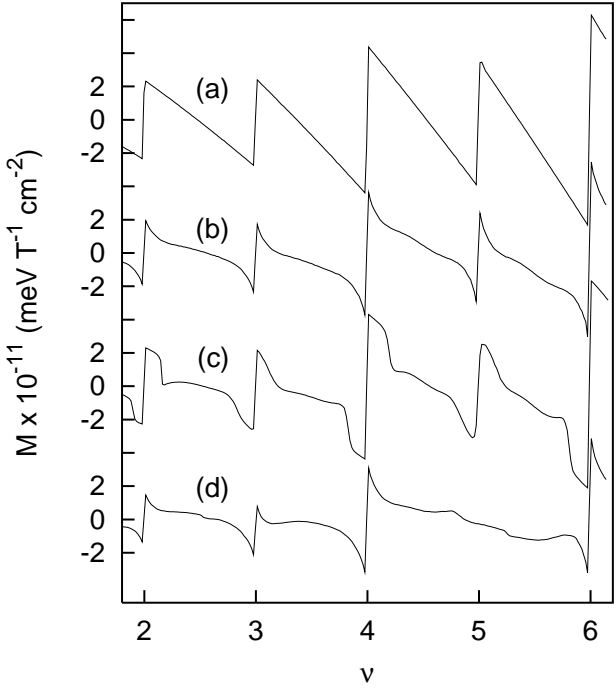}
\end{center}
\caption{ (a) The sawtooth magnetization for the homogeneous 2DEG in the
         HFA, with exchange-enhanced spin splitting. $B=3$ T.  
         (b) The effect of a one-dimensional modulation with $V_0=1.5$
         meV. (c) The effect of a disorder broadening $\Gamma=2.6$ meV 
         for the same modulation amplitude.
         (d) A modulation with $V=5$ meV, suppressing the exchange 
         enhancement of the spin splitting for $\nu=5$ ($\Gamma=0$).}
\label{1dmod}
\end{figure}

We have noticed in our calculations that for the system sizes 
considered here the
magnetization of the finite system is not strongly dependent on 
whether the modulation is assumed 2D as here, 
or 1D.\cite{PhysicaE} 
For an infinite system with a     
1D modulation we can calculate the thermodynamically defined
magnetization (\ref{thermodeq}) presented in Fig.\ \ref{1dmod}.
As mentioned before we see here that $M_o$
for the finite system, especially when it is enlarged, bears a strong
similarity with $M_o$ for the infinite 1D modulated 2DEG. 

The calculation of the ground state for a modulated 2DEG with
arbitrary magnetic fields is impossible for the 2D potential, due to
the commensurability restrictions.  The problem can be circumvented
for a 1D modulated system that we shall now turn our attention to,
and formulate predictions of experimental results that can be used to
test the importance of the exchange interaction.  For such a system we
have access to the total magnetization, i.\ e.\ bulk plus edge, in the
thermodynamic limit.  In Fig.\ \ref{1dmod} we display results for the
infinite system with a 1D potential, obtained with Eq. (\ref{thermodeq}).

First we show the sawtooth profile in the absence of a modulation
potential, reflecting the instability of the Fermi level in the energy
gaps.  The exchange interaction determines the spin splitting for odd
filling factors, but also amplifies the jumps for even filling factors by
almost a factor of two.  The reason is the enhancement effect on both the
spin gaps and the Landau gaps.   For the same reason, in the presence of
a modulation the jumps are only slightly reduced.  Similarily, as for the
2D modulation, the exchange interaction also increases the energy dispersion
of the Landau bands for noninteger filling factors, by lowering the
energy of the occupied states, see Figs.\ \ref{Ebpq1} and \ref{Ebpq2}.
Hence, the band width depends on the position of the Fermi level inside
an energy band.  This fact prevents the coincidence of the Fermi level
with a band edge (top or bottom), resulting in sharp cusps for $\nu$
close to integers. The sharpness is an effect of the exchange
interaction in the vicinity of a van Hove singularity.

Some amount of disorder may indeed broaden such cusps.  In addition, the
magnetization jumps may now slightly increase, because of the smearing
of the band edges by disorder, which helps the Fermi level to enter
or to leave a Landau band.  When the Fermi level lies in the middle
of a band screening effects are important.  In principle screening is
important when the modulation period is much bigger than the magnetic
length and/or when high Landau bands, with extended wave functions,
are occupied.  However, even here we can see some oscillations in the
upper bands, with orbital quantum number $n=2$.

Increasing the modulation amplitude, from $V_0=1.5$ meV to $V_0=5$
meV we first see that the magnetization for the bands with $n=1$
is relatively stable.  Just like in Figs.\ \ref{E_tot_cos_pq2} and
\ref{M_cos_pq2}, this is because the exchange amplification of the energy
dispersion depends on the filling factor, rather than on the modulation
amplitude.\cite{Manolescu99:5426}  The spin splitting survives now only
for $\nu=3$, and it is abruptly supressed for $\nu\geq5$.  A similar
supression occurs for $V_0=1.5$ meV, but at a higher $\nu$, and it can
be explained by the inflation of the wave functions in high Landau levels
which rapidly equilibrates the number of spin-up and spin-down electrons
and destroys the enhancement of the spin gap. \cite{Manolescu95:1703}
Such a suppression effect has been recently observed in magnetotransport
experiments on short-period modulated systems, in the Shubnikov - de
Haas peaks.\cite{Petit97:225}
%
%----------------------------------------------------------------
%
\section{Final remarks}
We have calculated the magnetization of periodic systems and 
discussed with examples various properties of it due to system boundaries,
periodicity, and Coulomb interaction.  We have compared the results
of the finite and infinite systems and of the periodic systems in one
and two spatial directions.  Our aim is to provide information for
understanding the magnetization measurements in mesoscopic systems, which
are expected to become a new direction of experimental investigations.

Unlike in other types of experiments, like transport or electromagnetic
absorption, the magnetization measurements seem to open a better and
more direct access to the intrinsic, quantum electronic structure of
the system.  In transport experiments this is often intermediated
by complicated electron-impurity interactions, and in far-infrared
absorption usually the classical collective motion of the electron
system is dominant.  We have identified in the magnetization several
properties of the energy spectrum which are absent or incompletely
observed in transport or absorption measuremens, like the exchange
enhancement of the energy dispersion or the curvature of the Landau
bands.  According to a recent prediction the exchange effects may also
determine hysteresis properties when acting on the energy dispersion,
either by varying the modulation amplitude, or by varying the Zeeman
splitting in tilted magnetic fields, and keeping the filling factor
constant.\cite{Manolescu99:5426}  The magnetization measurements could
be the best suited tool for probing such effects.
The present calculation further indicates that sought after delicate
internal structure of the Landau bands, such as the Hofstadter
butterfly,\cite{Hofstadter76:2239,Gudmundsson95:16744,Gudmundsson96:5223R}
in a doubly periodic 2DEG might be better observed by magnetization than
transport experiments.\cite{Schloesser96:683}
%
%----------------------------------------------------------------
%
\acknowledgments
This research was supported by the Icelandic Natural Science Foundation
and the University of Iceland Research Fund. 
In particular we acknowledge additional support from 
the Student Research Fund (S.\ E.), 
NATO Science Fellowship (A.\ M.), and 
Graduiertenkolleg "Physik Nanostrukturierter Festk\"orper" (V.\ G.).
%
%----------------------------------------------------------------
%
\bibliographystyle{prsty}
%\bibliography{/home/vidar/BIBtex/mod_qd}

%
%-----------------------------------------------------------------
%
%----------------------------------------------------------------
%
%
%----------------------------------------------------------------
%
%
%----------------------------------------------------------------
%
%
%----------------------------------------------------------------
%
%
%----------------------------------------------------------------
%
%
%-----------------------------------------------------------------
%
%
%------------------------------------------------------------------
%
%
%------------------------------------------------------------------
%
%
%\newpage
%-----------------------------------------------------------------
%
%
%-----------------------------------------------------------------
%
\newpage
%-----------------------------------------------------------------
%
%
%
%-----------------------------------------------------------------
%
%
%------------------------------------------------------------------
%
%
%-------------------------------------------------------------------
%
%
%--------------------------------------------------------------------
%
%-------------------------------------------------------------------
\end{document}